\documentclass[pra,aps,twocolumn,epsfig,superscriptaddress]{revtex4}
\usepackage{amsmath}
\usepackage{amssymb}
\usepackage[dvips]{graphicx}
\usepackage{dcolumn}
\usepackage{bm}
\usepackage[colorlinks=false,dvipdfm
]{hyperref} 
\usepackage{setspace}
\usepackage{amsfonts}
\usepackage{rotating}
\usepackage{makeidx}
\usepackage{amsthm}
\usepackage{bbold}
\usepackage{float}

\newcommand{\tr}{\mbox{Tr}}

\newcommand{\dd}{\mbox{d}}

\begin{document}

\title{Maxwell demon with  anti-decoherence}

\author{Zi-Yan Zhang}
\affiliation{Department of Physics, School of Science, Tianjin University, Tianjin 300072, China}

\author{Jian-Ying Du}
\affiliation{Department of Materials Science and Engineering, Southern University of Science and Technology,  Shenzhen 518055, China}

\author{Fu-Lin Zhang}
\email[Corresponding author: ]{flzhang@tju.edu.cn}
\affiliation{Department of Physics, School of Science, Tianjin University, Tianjin 300072, China}

\date{\today}

\begin{abstract}
Subsystems of a composite system in a pure state generally exist in mixed states and undergo changes with the overall state. This phenomenon arises from the coherence of the entire system and represents a crucial distinction between quantum and classical systems. Such a quantum property can enhance the work of an Otto heat engine, where two coupled qubits serve as the working substance, allowing situations in which negative work output initially occurred to now yield positive work. We utilize the imagery of Maxwell's demon to explain the reason for positive work in this Otto cycle, attributing it to the increased coherence after the mutual measurement of the two subsystems. Conversely, the quantum measurement-erase cycle typically outputs negative work, attributed to the decoherence of the instrument during the measurement process.
\end{abstract}

\keywords{}
	
 \maketitle
	
\section{Introduction}	

Quantum thermodynamics \cite{Book2004,Book2009} aims to establish thermodynamic laws on the foundation of quantum mechanics
and
investigates the impact of various quantum properties on  thermodynamic tasks.
Quantum information science offers  information-theoretic descriptions of various characteristics of quantum systems \cite{Nielsen,RMP2012Vedral},
establishing a natural connection between quantum thermodynamics and quantum information.
Actually, the discovery of the intimate relationship between information and thermodynamics  can be traced back to the research on Maxwell demon in 1871  \cite{MaxwellD} and the Szil\'{a}rd engine in 1929 \cite{Z.Phys.53.840}.
Many quantum versions of the  Maxwell demon and Szil\'{a}rd engine have been presented,
to investigate
 the  interplay between quantum information and thermodynamics \cite{RMP2009MaxwellD,discorddemons,NJP2017Szilard,PRL2017Demon,PRL2019NonequilibriumDemon,PRL2019,PRL2022Ren,PRL2011QSzilard,PRL2013Engine,NC2015,PRL.124.100603}.
The definitions of these models rely on the division between the quantum and classical worlds.
For instance,
in a recent classification  of Beyer \emph{et al.} \cite{PRL2019},
a \emph{truly quantum  demon}  is   one  obtaining system information
 through the quantum steering between the system and its environment.

Quantum correlations and measurement are two closely related  fundamental pillars of quantum information.
Their roles in thermodynamics have garnered extensive attention and investigation.
The correlations can be used to  enhance the extractable work
\cite{PhysRevX.5.041011,PhysRevE.93.052140,PhysRevA.99.052320,francica2017daemonic,PhysRevLett.121.120602,PhysRevLett.122.130601},
while their preparation is subject to the constraints of thermodynamic laws  \cite{Huber_2015,PhysRevE.100.012147}.
Similarly, measurements on quantum systems can fuel thermodynamic tasks
\cite{PRL2019MCool,PhysRevResearch.5.033122,PhysRevE.107.054110},
while the measurement-erase cycle also incurs the fundamental lower bounds on the thermodynamic energy cost \cite{PRL2009cost}.

The motivation for this work stems from exploring the thermodynamic effects of a simple characteristic of quantum correlations.
Specifically, when a composite system is in a pure state, its subsystems are generally in mixed states,
and these mixed states change with the evolution of the pure state.
This implies that when the whole system undergoes unitary evolution,
its subsystems may accomplish thermodynamic tasks that their classical counterparts cannot achieve.

In the present work,
 we focus on the ability of subsystems to output work during the quantum adiabatic processes of the whole systems.
These adiabatic processes are unitary, controlled by slowly varying local Hamiltonians, with constant interactions.
The eigenvalues of the whole density matrix remain unchanged, while the eigenstates remain consistent with the total Hamiltonian.
To investigate the ongoing effects of work generated due to the evolution of reduced states,
we consider an quantum Otto cycle composed of two adiabatic processes and two isochoric (thermalization) processes \cite{geva1992quantum,kieu2004second,quan2007quantum,thomas2011coupled}.
We choose a pair of coupled qubits as the working substance, and their local Hamiltonians vary in opposite ways.
This results in negative work output over a significant parameter range
 in the absence of interaction between them.
It is found that, in the presence of interactions during the two adiabatic processes,
the evolution of both subsystems contributes to positive work output.


The two qubits can be viewed as Maxwell demons measuring and controlling each other,
providing a localized perspective on their performance in the task of performing work.
The two adiabatic processes involve mutually acquiring information from each other and returning information, respectively.
Furthermore, in both processes, the occupation numbers of the qubits are controlled by each other,
evolving in a direction favorable for the generation of positive work.
Here, our \emph{measurement} has two characteristics,
distinguishing it from recent  definition  in the field of quantum thermodynamics \cite{PRL2019MCool,PhysRevResearch.5.033122,PhysRevE.107.054110}.
(i) Our measurement is driven by changes in the local Hamiltonians, while keeping the interactions constant;
typically studied measurements involve the turning on and turning off of interactions with the local Hamiltonians remaining unchanged.
(ii) After the two demons become entangled, they do not decohere and transition into a classical state.
 If demons undergo decoherence, followed by a restoration to the thermal equilibrium state before measurement, the entire cycle can only output negative work. However, our isochoric process following the measurement can increase coherence (i.e. it is an anti-decoherence process), which is precisely the key to the positive work output for the Otto cycle.

\section{Preliminaries}\label{PRE}

\subsection{Adiabatic Processes}

Let us begin with a general discussion of the adiabatic processes studied in this work.
The Hamiltonian of an $N$-body system is given by
 	\begin{equation}
     \mathcal{H}=\sum_{i=1}^{N} \mathcal{H}^{(i)}  +\mathcal{H}_I,
	\end{equation}
where $\mathcal{H}^{(i)}  $ is the local Hamitonian of the $i$th subsystem with  an external parameter $\omega^{(i)}$,
and $\mathcal{H}_I$ represents the invariant interaction between subsystems.
The system is prepared as a mixed state, $\rho$, composed of convex combinations of  eigenstates of $\mathcal{H}$.
Suppose the parameter $\omega^{(i)} $ varies slowly enough between $\omega^{(i)}_1 $  and $\omega^{(i)}_2 $,
causing $\mathcal{H}$ to vary between $\mathcal{H}_1$ and $\mathcal{H}_2$
, and no level crossings occur.
The quantum adiabatic theorem ensures that the occupancy of each eigenstate remains unchanged.
The work output by the whole system can be well defined as
	\begin{equation}\label{WG}
		W=-\int_{\mathcal{H}_1}^{\mathcal{H}_2}  \tr( \rho  \dd{\mathcal{H}}).
	\end{equation}
It can be expressed as the sum of contributions from each subsystem
	\begin{equation}\label{WL}
		W=-\sum_{i=1}^{N}  \int_{\omega^{(i)}_1}^{\omega^{(i)}_2}   \tr\biggr[ \rho^{(i)}\frac{\partial   \mathcal{H}^{(i)}} {\partial \omega^{(i)} } \biggr]  \dd \omega^{(i)},
	\end{equation}
where $\rho^{(i)}$ is the reduced state of $i$th subsystem,
and
 $\tr\left[ \rho^{(i)}\frac{\partial   \mathcal{H}^{(i)} } {\partial \omega^{(i)} } \right]$ can be regarded as the generalized force of the external parameter on it.

The eigenstates of the overall system are often coherent superpositions of local eigenstates, and they change with the local Hamiltonian.
This leads to variations in the occupancy of the subsystems as the overall system evolves adiabatically.
We restrict our discussion to the contribution of subsystems non-adiabaticity to the work.
When $[\sum_{i=1}^{N} \mathcal{H}^{(i)} ,\mathcal{H}_I]=0$, all eigenstates coincide with the local eigenstates.
That is, a non-diagonal (quantum) interaction in the local energy eigenbasis is required in order to generate the local non-adiabaticity.

\subsection{Otto Cycle}\label{freequbit}

To investigate the ongoing effects of non-adiabaticity in reduced states,
we consider a quantum Otto cycle with a pair of coupled qubits as the working substance in the next section.
Here, we first present a simple analysis of the case without coupling.

The Hamiltonian is represented by
	\begin{equation}\label{HL}
 \mathcal{H}_L=\omega^{A}(\openone^{A}+\sigma_z^{A})+\omega^{B}(\openone^{B}+\sigma_z^{B}),
 	\end{equation}
where $\sigma_z^{A/B}$ and $\openone^{A/B}$ are the third Pauli operator and identity operator of qubit $A/B$.
The existence of the identity operators ensures that the local ground energies are zero, simplifying the following analysis.
However, the final conclusions and graphical analysis remain unaffected by the presence or absence of identity operators.

\begin{figure}
 \centering
 \includegraphics[width=5 cm]{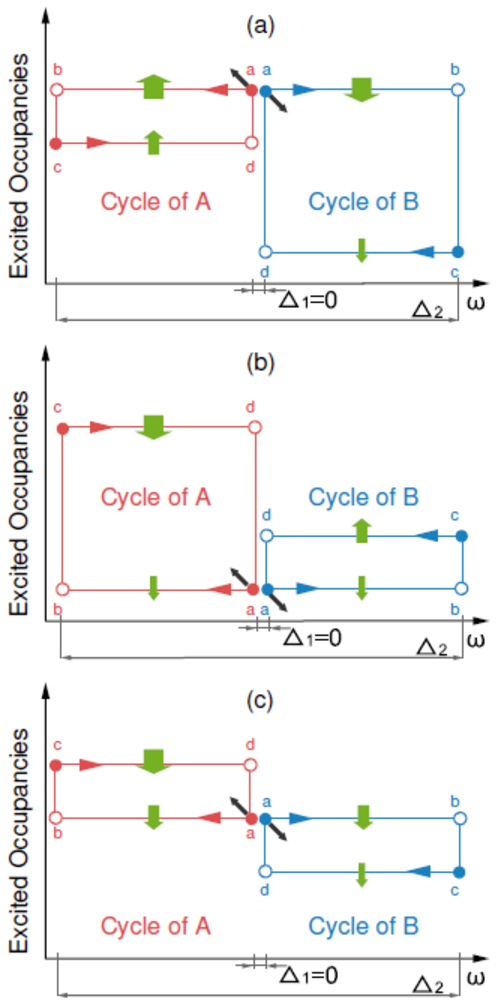}
 \caption{
(Color online)
Two uncoupled qubits exhibit only three possible scenarios for the occupancy of excited states during the Otto cycle with  $\Delta_1=0$, shown by (a), (b), and (c).
Qubit $A$  in (a) and qubit $B$ in (b) output positive work;
however, these works are cancelled out by the negative works of their counterparts.
To achieve positive work output, one can increase $\Delta_1$, causing the points corresponding to $\rho_a$ to change in the directions of the black arrows.
Filled circles represent thermal states, and empty ones represent  the nonequilibrium states at the ends of the adiabatic strokes.
}\label{Otto0}
\end{figure}

The external parameters (magnetic fields) can be  expressed in terms of their sum and difference as
	\begin{equation}
\Omega=\omega^{B}+ \omega^{A},\ \ \ \Delta=\omega^{B} - \omega^{A}.
 	\end{equation}
We are interested in cycles where $\omega^{A}$ and $\omega^{B}$ are varied oppositely.
Namely, for a fixed $\Omega$,  $\Delta$ varies back or forth between $\Delta_1$ and $\Delta_2$ with $0\leq\Delta_1<\Delta_2<\Omega$.
The complete cycle is outlined below, and Fig. \ref{Otto0} illustrates the scenario where $\Delta_1=0$ using the occupancies of local excited states.

\emph{\textbf{Adiabatic stroke: $a\rightarrow b$.}}
The working substance starts in a thermal state
$\rho_a=\exp( {-\beta_1 \mathcal{H}_{L1}})/\mathcal{Z}_1 $ with $\mathcal{H}_{L1}=\mathcal{H}_{L}(\Omega,\Delta_1)$ and $\mathcal{Z}_1 =\tr e^{-\beta_1 \mathcal{H}_{L1}}$.
That is, the two qubits equilibrate  with the first external thermal reservoir at temperature $T_1=1/\beta_1$ ($k_B=1$).
Then, they are isolated from the reservoir while $\mathcal{H}_{L}$ is varied into  $\mathcal{H}_{L2}=\mathcal{H}_{L}(\Omega,\Delta_2)$
sufficiently slowly.
Simultaneously, the state adiabatically evolves from $\rho_a$ to $\rho_b$.

\emph{\textbf{Isochoric stroke: $b\rightarrow c$.}}
The two qubits  are put in
contact with the second external thermal reservoir at temperature $T_2=1/\beta_2$ and allowed to relax with a fixed Hamiltonian $\mathcal{H}_{L2}$
until they reach the thermal state $\rho_c=\exp( {-\beta_2 \mathcal{H}_{L2}})/\mathcal{Z}_2 $ with  $\mathcal{Z}_2 =\tr e^{-\beta_2 \mathcal{H}_{L2}}$.

\emph{\textbf{Adiabatic stroke: $c\rightarrow d$.}}
The system is removed from contact with
the second thermal reservoir, and the Hamiltonnian is reversed back to $\mathcal{H}_{L2}$.
Simultaneously, the state adiabatically evolves from $\rho_c$ to $\rho_d$.

\emph{\textbf{Isochoric stroke: $d\rightarrow a$.}}
The  two qubits come into contact with the first thermal reservoir again,
and are thermalized to the equilibrium state $\rho_a$.

Only the two adiabatic processes produce work.
The work done by each subsystem can be calculated using Eq. (\ref{WL}).
The local Hamiltonians $\mathcal{H}_A=2 \omega^{A} |0\rangle_A \langle 0|$ and  $\mathcal{H}_B=2 \omega^{B} |0\rangle_B \langle 0|$,
result in the work of $A$ and $B$
	\begin{equation}
W_A= -2 \oint p^A d \omega^A,\ \ \  W_B= -2 \oint p^B d \omega^B,
 	\end{equation}
where $p^A$ and $p^B$ are
the occupancies of local excited states of $A$ and $B$, and the integrals are taken along the cycle.
They are proportional to the counterclockwise area traced out by the respective subsystems in the excited occupancies-$\omega$ space (as shown in Fig. \ref{Otto0}) , while a clockwise cycle yields negative work.

A deterministic conclusion here is that, when $\Delta_1 = 0$, the total work output of such an Otto cycle is negative, i.e., $W = W_A + W_B < 0$,
regardless of the values of the two temperatures, $\Omega$, or $\Delta_2$.
Since the two subsystems are non-interacting, they each undergo an Otto cycle independently, forming two rectangles of equal width in the excited occupancies-$\omega$ space.
In the equilibrium state $\rho_a$, $\omega^A = \omega^B$, thus $p^A_a = p^B_a$; in state $\rho_c$, $\omega^A < \omega^B$, so $p^A_c > p^B_c$.
This results in only three possible relationships between these two rectangles as shown in Fig. \ref{Otto0}, leading to a negative total work.

To obtain positive work, one can change the value of $\Delta_1$ and choose appropriate temperatures and $\Omega$ such that
\begin{equation}\label{Wplus0}
p^A_a - p^A_c > p^B_a - p^B_c.
 \end{equation}
We will not delve into this approach further.
In the next section, we will demonstrate that in the presence of interactions, the non-adiabaticity of local density matrices during the overall adiabatic processes can also lead to similar results.

\section{Coupled Cycle}\label{Otto}

To build coherence of the entire system, we introduce a two-qubit interaction
	\begin{equation}%
\mathcal{H}_I=J\left(\sigma_+^{A}\sigma_-^{B}+\sigma_-^{A}\sigma_+^{B}\right)
		\label{}
	\end{equation}
where $\sigma_+^{\alpha}=|0\rangle_{\alpha} \langle 1|$ and $\sigma_-^{\alpha}=|1\rangle_{\alpha} \langle 0|$
are the raising and lowering operators for  qubits $\alpha=A$ and $B$, and $J$ is the interaction strength.
The eigenvectors and the corresponding eigenvalues of the total
Hamiltonian
	\begin{equation}
\mathcal{H}=\mathcal{H}_L+\mathcal{H}_I
		\label{HXX}
	\end{equation}
 are given by
 	\begin{equation}\label{eigen}
		\begin{aligned}
&|\phi_0\rangle=|00\rangle_{AB},  &&E_0=2\Omega; \\
&|\phi_{+}\rangle=\cos{\theta}|10\rangle_{AB}+\sin{\theta}|01\rangle_{AB}, &&E_{+}= \Omega + D; \\
&|\phi_{-}\rangle=\cos{\theta}|01\rangle_{AB}-\sin{\theta}|10\rangle_{AB}, &&E_{-}= \Omega - D;  \\
&|\phi_1\rangle=|11\rangle_{AB},   &&E_1=0;
	\end{aligned}
	\end{equation}
where $D=\sqrt{\Delta^2+J^2}$, and $\theta$ is defined by $\sin{2\theta}=J/D$ and $\cos{2\theta}=\Delta/D$.

In this section, we consider an Otto cycle identical to the case of free qubits, with only a slight modification in the parameter constraints: $0\leq\Delta_1<\Delta_2<\sqrt{\Omega^2-J^2}$ to ensure  no level crossing.
The discussions in subsection \ref{freequbit} before the conclusion (i.e., when $\Delta_1=0$, $W<0$) remain applicable here, with the only difference being the replacement of $\mathcal{H}_L$ with the total Hamiltonian $\mathcal{H}$.
The differential in Eq. (\ref{WG}) is given by
\begin{equation}
		\begin{aligned}
 \dd \mathcal{H} = & (|\phi_{+}\rangle \langle \phi_{+}| - |\phi_{-}\rangle \langle \phi_{-}|) \dd D \\
 & + E_+ \dd (|\phi_{+}\rangle \langle \phi_{+}| ) + E_-  \dd (|\phi_{-}\rangle \langle \phi_{-}|),
		\end{aligned}
\end{equation}
and the reduced states of $|\phi_{\pm}\rangle$
vary during the adiabatic processes.
Consequently, both the total work and the contribution of each subsystem primarily depend on the properties of the two entangled states $|\phi_{\pm}\rangle$.

The Hamiltonian (\ref{HXX}) can be diagonalized by a global unitary transformation into two free \emph{pseudo} qubits,
which are the result of entanglement between the two physical qubits \cite{Du_2018}.
The
thermodynamic functionalities of  our current model  can be understood in the view of these two pseudo qubits,
utilizing the results provided in subsection \ref{freequbit}.
We bypass this imagery here, and instead understand it from the perspective of the system as a whole and the two physical qubits separately.
Thus, our qualitative conclusions are not limited to the Heisenberg isotropic XX type Hamiltonian (\ref{HXX}) but are generally applicable to models of two coupled qubits.

\subsection{Global Analysis}\label{global }

The total work of the cycle,
 contributed by the two adiabatic processes,
 can be directly calculated using Eq. (\ref{WG}) as
	\begin{equation}\label{Wtotal}
		W=\left[( p_{a}^{-} -p_{a}^{+}  ) -  ( p_{c}^{-} -p_{c}^{+}  )\right] (D_2-D_1 ),
	\end{equation}
where $D_{1,2} = \sqrt{\Delta_{1,2}^2 +J^2}$,
and $p_{a,c}^{\pm}$  are the populations of $|\phi_{\pm}\rangle$ in the thermal states $\rho_{a,c}$.
Under the condition $0\leq\Delta_1<\Delta_2<\sqrt{\Omega^2-J^2}$,
, the positive work window is determined by
\begin{equation}\label{Wplus}
       p_{a}^{-} -p_{a}^{+}    >    p_{c}^{-} -p_{c}^{+} .
\end{equation}
When $J=0$, $p_{a,c}^{+} =p^B_{a,c}(1-p^A_{a,c}) $ and $p_{a,c}^{-} =p^A_{a,c}(1-p^B_{a,c})$,
and the window returns into the inequality (\ref{Wplus0}).
The heat absorbed by the system during the two isochoric processes can be calculated by the increase in its energy as
 $Q_1=\tr [(\rho_a-\rho_d) \mathcal{H}_1]$ and $Q_2=\tr [(\rho_c-\rho_b) \mathcal{H}_2]$ with $\mathcal{H}_{1,2}=\mathcal{H}_{L1,2} +\mathcal{H}_{I}$.
They are
\begin{equation}
\begin{aligned}
&Q_1= (\delta p^0- \delta p^1)\Omega- (\delta p^- - \delta p^+ )D_1,\\
&Q_2=-(\delta p^0- \delta p^1)\Omega+ (\delta p^- - \delta p^+ )D_2,
\end{aligned}
\label{Q1Q2}
\end{equation}
where $\delta p^{0,1,\pm}=  p_a^{0,1,\pm}-p_c^{0,1,\pm}$
 represents the difference in the occupancies of the corresponding eigenstates $|\phi_{0,1,\pm}\rangle$
 between the two thermal states $\rho_a$ and $\rho_c$.

\begin{figure}
 \includegraphics[width=2.7cm]{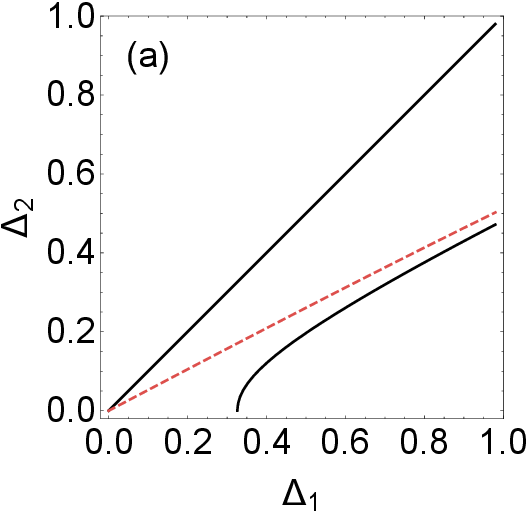}
 \includegraphics[width=2.7cm]{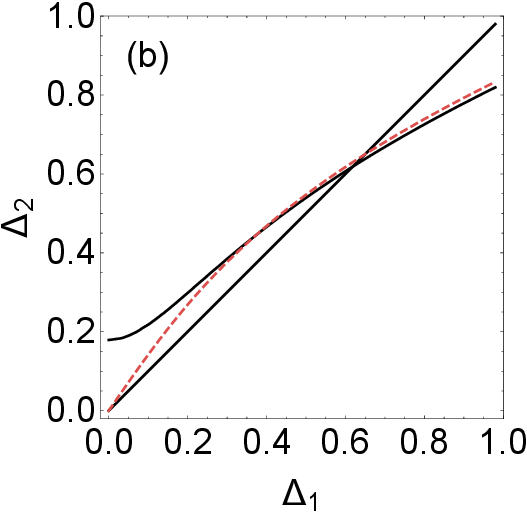}
 \includegraphics[width=2.7cm]{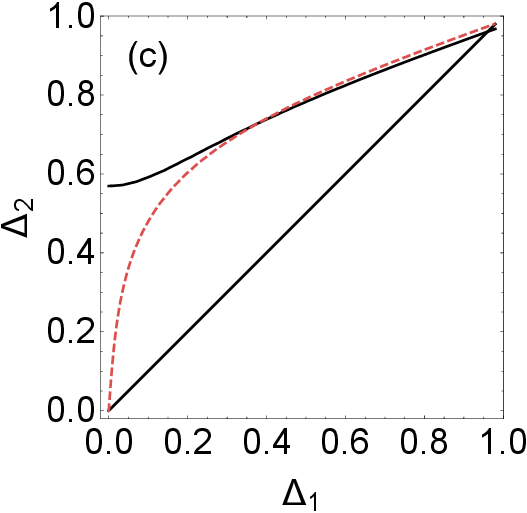}
 \caption{
(Color online)
The two solid lines in each plot correspond to $\Delta_2=\Delta_1$ and $f=0$, for fixed $\beta_2/\beta_1 =3$,  $\Omega=1$, and $J=0.2$.
The region between them represents the positive work window.
In (a), (b), and (c), $\beta_1=0.5,1,2$ respectively.
The dashed line represents the result of $f=0$ when $J=0$.
}\label{Workfig}
\end{figure}

The positive work window (\ref{Wplus}) under the condition $0\leq\Delta_1<\Delta_2<\sqrt{\Omega^2-J^2}$ may require either $T_1 > T_2$ or $T_1 < T_2$, depending on the values of $\Delta_1$, $\Delta_2$, $\Omega$, and $J$.
For the sake of convenience in discussion, in the remaining part of this subsection, we assume $T_1 > T_2$, and we relax the constraints on the external parameters to $\Delta_1,\Delta_2\in \left[ 0,\sqrt{\Omega^2-J^2}\right)$, which is without loss of generality.
The positive work window can be classified based on the properties of $f = ( p_{a}^{-} -p_{a}^{+} ) - ( p_{c}^{-} -p_{c}^{+} )$ and $D_2-D_1$:
the former is a monotonically decreasing function of $\Delta_2$, while the latter increases with $\Delta_2$.
As illustrated in Fig. \ref{Workfig}, in the ($\Delta_1$, $\Delta_2$) space, the positive work region is bounded by the zeros of these two functions.
 When $\Delta_2 = \Delta_1$, if $f < 0$, positive work occurs in $\Delta_1 > \Delta_2$;
 conversely, if $f > 0$, positive work occurs in $\Delta_1 < \Delta_2$.
 Given the ratio of temperatures $T_2 / T_1$, the former occurs in the high-temperature and larger $\Delta_1$ region, while the latter occurs in the low-temperature and smaller $\Delta_1$ region.

\begin{figure}
 \includegraphics[width=7.1cm]{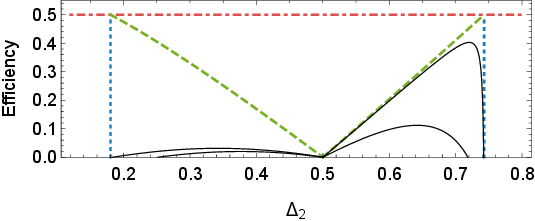}
 \caption{
 (Color online)
Solid lines show efficiency, with their heights corresponding to parameters $\beta_1=5,2.5,0.1,0.5$ in decreasing order, for fixed $\beta_2/\beta_1 =2$, $\Delta_1=0.5$, $\Omega=1$ and $J=0.2$.
The dash-dotted horizontal line represents the Carnot efficiency,
 while the dashed lines represent the upper bound of efficiency $\eta_{up}$.
 The two vertical dotted lines indicate the maximum range of $\Delta_2$ for positive work.
}\label{eta}
\end{figure}

The properties of the positive work window described above can be understood using two extreme cases:
 the high-temperature limit and the low-temperature limit.
For finite values of $x=\Omega,D_1,D_2$, when $\beta \rightarrow +\infty$ (low-temperature limit), $e^{\beta x} \pm e^{-\beta x} \rightarrow e^{\beta x}$,
and consequently $f\rightarrow e^{-\beta_1 (\Omega-D_1)}-e^{-\beta_2 (\Omega-D_2)}$.
In this case, only the lowest two energy levels exist,
which constitute a qubit Otto heat engine with an energy gap of $\Omega-D$.
Substituting $f$ into Eq. (\ref{Wtotal}), one obtains the condition for positive work as
\begin{equation}\label{Dleft}
\beta_2 (\Omega-D_2)>\beta_1 (\Omega-D_1).
\end{equation}
When $\beta \rightarrow  0$ (high-temperature limit), $e^{\beta x} \rightarrow 1+\beta x $, and thus $f\rightarrow \frac{1}{2}(\beta_1 D_1 -\beta_2 D_2)$.
In this case, the total probability of the middle two energy levels $|\phi_{\pm}\rangle$ remains $1/2$ throughout the cycle,
constituting a qubit Otto heat engine with an energy gap of $2D$.
Similarly, we obtain the condition for positive work as
\begin{equation}\label{Dright}
\beta_2  D_2 >\beta_1 D_1.
\end{equation}
These two inequalities provide the maximum possible positive work region when $T_2 / T_1$ is fixed.
The involvement of other levels outside of these two \emph{effective} Otto heat engines will counteract the work or absorb more heat,
thus providing an upper limit on the efficiency of the entire system (as shown in Fig. \ref{eta}), i.e.,
\begin{equation}\label{etabound}
\eta=\frac{W}{Q_1}
< \eta_{up} =\biggr \{
\begin{aligned}
&1-\frac{D_2}{D_1},  \ \ \  & \Delta_2<\Delta_1 ;\\
&1-\frac{\Omega-D_2}{\Omega-D_1},   \ \ \  & \Delta_2>\Delta_1 .
\end{aligned}
\end{equation}
Clearly, this upper bound is lower than the Carnot efficiency, $\eta_{up} < \eta_{c} = 1 - T_2/T_1$,
thus complying with thermodynamic principles.
The proof
 for the maximum possible positive work region and the relations (\ref{etabound})
can be found in Appendix \ref{WindowAndEta} .

\subsection{Local View}\label{Demon}

\begin{figure}
 \centering
 \includegraphics[width=5cm]{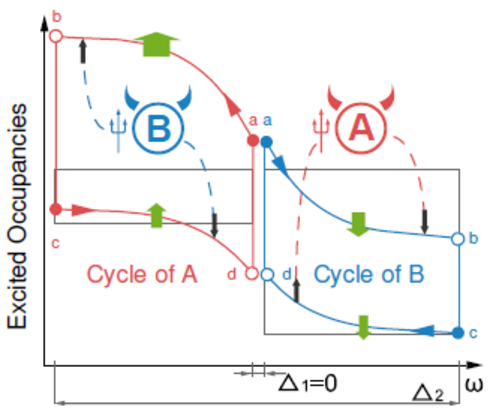}
 \caption{
  (Color online)
Two coupled qubits act as Maxwell demons measuring and controlling each other.
In both adiabatic strokes,
the occupation numbers (colored lines) of the qubits are evolved in a direction favorable for the generation of positive work, as shown by  the black arrows.
Except for a nonzero $J$, the parameters are the same as in Fig. \ref{Otto0} (a),
which is shown as the grey lines.
The evolution trend of occupation numbers in the other two cases in Fig. \ref{Otto0} is the same as in this figure,
because the effect arises from $|\phi_{\pm}\rangle$ and the probability of $|\phi_{-}\rangle$ is higher than that of $|\phi_{+}\rangle$.
Filled circles represent the reduced states of thermal states, and empty ones represent  the ones of nonequilibrium states at the ends of the adiabatic strokes.
}\label{Demonfig}
\end{figure}

Now we return to the perspective of subsystems to consider the impact of the interaction on the positive work regions.
As shown in Fig. \ref{Workfig}, when $T_1>T_2$, the positive work regions where $\Delta_2<\Delta_1$ and $\Delta_2>\Delta_1$ and $\Delta_1$ is small are expanded due to interactions.
This result can be explained by the influence of interactions on the eigenvalues of the total Hamiltonian.
However, a local perspective allows us to observe the role of information.
 Just as in the classical Maxwell demon model, the demon and the system it measures and controls, along with the demon's memory,
 constitute two subsystems of a larger system.
The interactions contributes to the work of the subsystems in two ways:
(1) altering the reduced density matrices of the equilibrium states;
(2) inducing the evolution of subsystems during adiabatic processes.
We focus on the effect of the second point.

Here, we no longer require $T_1>T_2$, and restrict our discussion to the context of $\Delta_1=0$,
although the follow illustrative representations are more generally applicable.
In this scenario, as illustrated in Fig. \ref{Otto0}, without coupling, the entire system deterministically cannot output positive work.
However, when the interaction is present, such Otto cycles can generate positive work both for $T_1<T_2$ and $T_1>T_2$.
This is evident from the regions of positive work on the horizontal axis of Fig. \ref{Workfig} (a), with  interchanging the subscripts of $\Delta_1$ and $\Delta_2$, $T_1$ and $T_2$, as well as on the vertical axes of (b) and (c).
Additionally, the evolution of subsystems during adiabatic processes is necessary for generating positive work, as $p^A_a = p^B_a$ and $p^A_c > p^B_c$  [see Eq. (\ref{PAPB}) below] still hold in the two coupled equilibrium states $\rho_a$ and $\rho_c$.

The evolution of the two subsystems achieves positive work  that is unattainable without the interaction.
This kind of evolution relies on mutual measurement and control (i.e. establishing correlation and altering each other's states), making them analogous to Maxwell demons for each other.
One can derive the occupation numbers of the local excited states during the adiabatic processes as
\begin{equation}\label{PAPB}
\begin{aligned}
& p^A=p^0+\frac{1}{2}(p^- + p^+)+\frac{1}{2}(p^- - p^+) \cos 2\theta,\\
 &p^B=p^0+\frac{1}{2}(p^- + p^+)-\frac{1}{2}(p^- - p^+) \cos 2\theta,
\end{aligned}
\end{equation}
where $p^{0,\pm}$ are the populations of the global eigenstates $|\phi_{0,\pm}\rangle$.
As illustrated in Fig. \ref{Demonfig}, accompanying the adiabatic evolution of the overall system,
they are elevated when the corresponding subsystem performs positive work and depressed when it performs negative work.
Since $\omega^{A}$ and $\omega^{B}$ are varied oppositely,
the total work (\ref{Wtotal}) is determined by the parts of $p^A$ and $p^B$ that evolve during the adiabatic processes.

Although mutual control during the two adiabatic processes is beneficial for achieving positive work,
their significance in measurement is different.
When $J \ll \Delta_2$, $\rho_c \approx \rho^A_c \otimes \rho^B_c$ with  $\rho^A_c $ and $\rho^B_c$ being the two reduced states,
and $|\phi_{\pm}\rangle$ in $\rho_d$ and $\rho_a$ are two Bell states.
The  stroke from $c$ to $d$ represents the establishment of correlation between the two systems and the acquisition of information from each other,
while in the one  from  $a$ to $b$, they unitarily return information to each other.
The difference here from measurements in the usual literature is that our interactions remain unchanged,
while the establishment of correlations is controlled by local Hamiltonians.
Additionally, the information of the system is not acquired by external classical instruments and used for feedback control,
thus the two qubits do not decohere after establishing correlations.

As a comparison, one can envision a decoherence process from $d$ to $\tilde{a}$:
Firstly, adiabatically turn off the interaction, keeping the system state unchanged as $\rho_d$;
Secondly, as the two subsystems act as measuring devices, their off-diagonal elements vanish, leading to the system state becoming $\rho_{\tilde{a}}$.
Here, $\rho_{\tilde{a}}=\mathcal{D}(\rho_d)$, where $\mathcal{D}$ represents the dephasing operation in the local eigenstates.
If one solely utilize a heat source at temperature $T_2$ and external work to return the  system to $\rho_c$,
the process from $\rho_c \rightarrow \rho_d\rightarrow \rho_{\tilde{a}} \rightarrow\rho_c$ constitutes a measurement-erase cycle.
The total work of this cycle is given by [derivations are given in the Appendix \ref{WorkME}]
\begin{equation}\label{WorkDec}
\tilde{W} \leq - T_2  C_r(\rho_d)
\end{equation}
where $C_r(\rho_d)$ is the relative entropy of coherence of $\rho_d$ \cite{PhysRevLett.113.140401}.
$C_r(\rho_d)$ also quantifies the reduction of overall coherence from $d$ to $\tilde{a}$.
Therefore, we can expect that if the process from $d$ to $a$ enhances coherence (anti-decoherence), then the total work of the entire Otto cycle can be positive.
This provides an intuitive understanding of the positive work condition (\ref{Wplus}),
as it is equivalent to $C_{l_1}(\rho_a)>C_{l_1}(\rho_d)$,
where $C_{l_1}$ represents the $l_1$-norm of coherence \cite{PhysRevLett.113.140401}.

\section{summary}\label{summary}

 In summary, we investigated the contribution of subsystem evolution during adiabatic processes to work output using a pair of coupled qubits implementing an Otto cycle.
  In this model, each qubit serves as a Maxwell demon for the other, interacting to exchange information and steering their respective evolutions during adiabatic processes toward states favorable for work generation.
  This results in positive work output even in cases where it would otherwise be impossible.

The extension in three directions is particularly worthy of consideration.
Firstly, the role of subsystem evolution in other thermodynamic tasks is a natural question, such as refrigeration or driving other quantum machines.
Secondly,  a more general quantum information theory analysis of similar processes, where measurement feedback occurs within the system,
would lead to a better understanding of the relationship between quantum information and thermodynamics.
Additionally, when the evolution time of external parameters is finite,
the interaction between the overall system's nonadiabatic effects and the subsystem's nonadiabaticity is intriguing.

\begin{acknowledgments}
This work was supported by the National Natural Science Foundation of China (Grants No. 11675119 and No. 11575125).
\end{acknowledgments}

\begin{appendix}

\section{Maximum positive work region and  maximum efficiency}\label{WindowAndEta}

For a two-level system undergoing an Otto cycle between two heat baths at temperatures $T_1$ and $T_2$,
and with the energy gap changing between $2\omega_1$ and $2\omega_2$, the output work can be directly calculated as
\begin{equation}\label{Wdiwenjinsif}
W'=\left[\tanh(\beta_2\omega_2)-\tanh(\beta_1\omega_1)\right]\left(\omega_1-\omega_2\right).
\end{equation}
When $\beta_1<\beta_2$, $W'>0$ can only occur when $\omega_1>\omega_2$ and $\beta_1\omega_1<\beta_2\omega_2$.
The heat absorbed from the high-temperature reservoir is
\begin{equation}
Q_1'=\left[\tanh(\beta_2\omega_2)-\tanh(\beta_1\omega_1)\right]\omega_1.
\end{equation}
Thus, its efficiency is
\begin{equation}
\eta'=\frac{W'}{Q_1'}=1-\frac{\omega_2}{\omega_1}<1-\frac{\beta_1}{\beta_2}.
\end{equation}

\textbf{Case: $\Delta_2 > \Delta_1$.}
At this point, we take $\omega_1=(\Omega-D_1)/2$ and $\omega_2=(\Omega-D_2)/2$,
and compute their difference with the total work (\ref{Wtotal}) to obtain
\begin{equation}\label{DifferWWOTLow}
W-W'=\left(\omega'_2 -\omega'_1 \right)\left[\tanh(\beta_1\omega'_1 )-\tanh(\beta_2\omega'_2)\right],
\end{equation}
where $ \omega'_1=(\Omega+D_1)/2$ and $ \omega'_2=(\Omega+D_2)/2$.
Thus, when $\beta_1<\beta_2$ and $\Delta_2 > \Delta_1$, $W<W'$.
In other words, inequality (\ref{Dleft}) is a necessary condition for $W>0$ in the region of $\Delta_2 > \Delta_1$.
Simultaneously,
\begin{equation}
Q_1-Q_1'= \omega'_1 \left[\tanh(\beta_2\omega'_2 )-\tanh(\beta_1\omega'_1)\right]>0,
\end{equation}
which implies
\begin{equation}
\eta<\eta'=1-\frac{\Omega-D_2}{\Omega-D_1}.
\end{equation}

\textbf{Case:  $\Delta_2 < \Delta_1$.}
The total work (\ref{Wtotal}) can be expressed as
\begin{equation}
W=\left[P_2 \tanh(\beta_2 D_2)-P_1 \tanh(\beta_1 D_1)\right]\left(D_1-D_2\right),
\end{equation}
where $P_i= \cosh(\beta_i D_i)/[ \cosh(\beta_i D_i)+ \cosh(\beta_i \Omega)]$ with $i=1,2$.
When the system only consists of the middle two energy levels,
the work done by the Otto cycle is given by
\begin{equation}
W''=\left[ \tanh(\beta_2 D_2)- \tanh(\beta_1 D_1)\right]\left(D_1-D_2\right).
\end{equation}
When $\beta_1<\beta_2$, in the region where $W''\leq0$, i.e.  $\beta_1 D_1 \geq \beta_2 D_2 $,
we have
\begin{equation}
\frac{\cosh(\beta_1 \Omega)}{\cosh(\beta_1D_1)} < \frac{\cosh(\beta_2 \Omega)}{\cosh(\beta_2 D_2)}.
\end{equation}
This leads to: $P_2 < P_1 $.
The conclusion is, when $W''<0$, there must be $W<0$.
Therefore, the inequality (\ref{Dright}) servers as  a necessary condition for $W>0$ in the region where $\Delta_2 < \Delta_1$.

The heat in (\ref{Q1Q2}) absorbed from the high-temperature reservoir can be written as
\begin{equation}
\begin{aligned}
Q_1 =& \left[P_2 \tanh(\beta_2 D_2)-P_1 \tanh(\beta_1 D_1)\right] D_1\\
         & + (S_2- S_1)\Omega,
\end{aligned}
\end{equation}
where
\begin{equation}
S_i=\frac{\sinh (\beta_i \Omega)}{\cosh(\beta_i \Omega)+\cosh (\beta_i D_i) },
\end{equation}
with $i=1,2$.
For a fixed $ D_i\in[0,\Omega]$,
$S_i$  is a monotonically increasing function of $\beta_i$.
Therefore,
\begin{equation}
S_2 > \frac{\sinh (\beta_2 \Omega)}{\cosh(\beta_2 \Omega)+\cosh (\beta_2 D_1) } >S_1.
\end{equation}
Then, the  efficiency
\begin{equation}
\eta<\eta''=1-\frac{ D_2}{ D_1}.
\end{equation}

\section{Total work of measurement-erase cycle}\label{WorkME}

We calculate the work done by the system at each step sequentially.
Firstly, during the adiabatic evolution from state $c$ to $d$, the work done by the system is given by
\begin{equation}\label{Wcd}
W_{c\to d}=\tr (\rho_c  \mathcal{H}_2)-\tr (\rho_d  \mathcal{H}_1),
\end{equation}
where $ \mathcal{H}_1= \mathcal{H}_{L1}+\mathcal{H}_I $ and $ \mathcal{H}_2= \mathcal{H}_{L2}+\mathcal{H}_I $.
Secondly, during the process of turning off the interaction,
\begin{equation}\label{Wto}
W{\text{turn off}}= \tr (\rho_d  \mathcal{H}_1)-\tr (\rho_d \mathcal{H}_{L1} ),
\end{equation}
and furthermore, during the evolution from $\rho_d$ to $\rho_{\tilde{a}}$, where the system's Hamiltonian remains unchanged,
\begin{equation}\label{Wdc}
W_{\text{decoherence}}= 0.
\end{equation}

In the final step, the system returns from $\rho_{\tilde{a}}$ to the equilibrium state $\rho_c$, which is equivalent to erasing the information obtained by each subsystem from the other.
This process requires external control, involving the transition of $\mathcal{H}_{L1}$ back to $\mathcal{H}_{L2}$,
the turning-on of $\mathcal{H}_I$,
and thermal contact with the reservoir at temperature $T_2$.
The thermal equilibrium state of the reservoir is $\rho_r=\exp (-\beta_2 \mathcal{H}_r)/\mathcal{Z}_r$,
where $\mathcal{H}_r$ is the Hamiltonian of the reservoir, and $\mathcal{Z}_r=\tr e^{-\beta_2 \mathcal{H}_r}$ is the partition function.
The total unitary process experienced by the combined system of the system and the reservoir is given by
\begin{equation}
\mathcal{U} \rho_{\tilde{a}} \otimes \rho_r \mathcal{U}^{\dag} =\rho_{sr},
\end{equation}
where the reduced density matrix of the system in $\rho_{sr}$ is $\tr{r} \rho_{sr} =\rho_c$.
In this process, the work obtainable by the external agent is given by
\begin{equation}\label{Wac}
W_{\text{erase}}= \tr [\rho_{\tilde{a}} \otimes \rho_r (\mathcal{H}_{L1}+\mathcal{H}_{r})]-\tr [\rho_{sr} (\mathcal{H}_{2}+\mathcal{H}_{r})].
\end{equation}
Substituting the results of equilibrium states
 \begin{equation}
\begin{aligned}
\mathcal{H}_{2}+\mathcal{H}_{r} &=-T_2[ \ln (\rho_c \otimes \rho_r) + \ln \mathcal{Z}_2 + \ln  \mathcal{Z}_r],\\
\tr(\rho_c \mathcal{H}_2) &= T_2 \mathcal{S}(\rho_c) - T_2 \ln \mathcal{Z}_2,\\
\tr(\rho_r \mathcal{H}_r) &= T_2 \mathcal{S}(\rho_r) - T_2 \ln \mathcal{Z}_r,
\end{aligned}
\end{equation}
where  $\mathcal{Z}_2=\tr e^{-\beta_2 \mathcal{H}_{2}}$ and $\mathcal{S}$ denotes von Neumann entropy,
one can find that
\begin{equation}
\begin{aligned}
W_{\text{erase}}=&  T_2 \tr[\rho_{sr}  \ln (\rho_c \otimes \rho_r) ] - \tr (\rho_c \mathcal{H}_2) \\
&+ \tr(\rho_{\tilde{a}} H_{L1}) +   T_2 \mathcal{S}(\rho_c) + T_2 \mathcal{S}(\rho_r) .
\end{aligned}
\end{equation}
Finally, utilizing the Klein inequality and $\mathcal{S}(\rho_{\tilde{a}} \otimes \rho_r)=\mathcal{S}(\rho_{sr})$, we obtain
\begin{equation}
W_{\text{erase}} \leq  \tr(\rho_{\tilde{a}} H_{L1}) -  \tr (\rho_c \mathcal{H}_2)  +T_2 [\mathcal{S}(\rho_c)- \mathcal{S}(\rho_{\tilde{a}} )].
\end{equation}

We add up the work in these four steps, yielding
\begin{equation}
\tilde{W} \leq T_2 [\mathcal{S}(\rho_c)- \mathcal{S}(\rho_{\tilde{a}} )],
\end{equation}
where we have utilized $\tr(\rho_{\tilde{a}} H_{L1}) = \tr(\rho_{d} H_{L1})$,
since $\rho_{\tilde{a}} $ is the result of decoherence of $\rho_{d} $ in the local energy eigenbasis.
Furthermore, $\mathcal{S}(\rho_c)=\mathcal{S}(\rho_d)$ leads to
\begin{equation}
\tilde{W} \leq - T_2 [\mathcal{S}(\rho_d)- \mathcal{S}(\mathcal{D}(\rho_{d}) )],
\end{equation}
which is the inequality (\ref{WorkDec}).

\end{appendix}


\bibliography{CoupledOtto}

\begin{thebibliography}{35}
\expandafter\ifx\csname natexlab\endcsname\relax\def\natexlab#1{#1}\fi
\expandafter\ifx\csname bibnamefont\endcsname\relax
  \def\bibnamefont#1{#1}\fi
\expandafter\ifx\csname bibfnamefont\endcsname\relax
  \def\bibfnamefont#1{#1}\fi
\expandafter\ifx\csname citenamefont\endcsname\relax
  \def\citenamefont#1{#1}\fi
\expandafter\ifx\csname url\endcsname\relax
  \def\url#1{\texttt{#1}}\fi
\expandafter\ifx\csname urlprefix\endcsname\relax\def\urlprefix{URL }\fi
\providecommand{\bibinfo}[2]{#2}
\providecommand{\eprint}[2][]{\url{#2}}

\bibitem[{\citenamefont{Gemma et~al.}(2004)\citenamefont{Gemma, Michel, and
  Mahler}}]{Book2004}
\bibinfo{author}{\bibfnamefont{J.}~\bibnamefont{Gemma}},
  \bibinfo{author}{\bibfnamefont{M.}~\bibnamefont{Michel}}, \bibnamefont{and}
  \bibinfo{author}{\bibfnamefont{G.}~\bibnamefont{Mahler}},
  \emph{\bibinfo{title}{Quantum Thermodynamics}} (\bibinfo{publisher}{Springer,
  Berlin}, \bibinfo{year}{2004}).

\bibitem[{\citenamefont{Gemmer et~al.}(2009)\citenamefont{Gemmer, Michel, and
  Mahler}}]{Book2009}
\bibinfo{author}{\bibfnamefont{J.}~\bibnamefont{Gemmer}},
  \bibinfo{author}{\bibfnamefont{M.}~\bibnamefont{Michel}}, \bibnamefont{and}
  \bibinfo{author}{\bibfnamefont{G.}~\bibnamefont{Mahler}},
  \emph{\bibinfo{title}{Quantum Thermodynamics: Emergence of Thermodynamic
  Behavior Within Composite Quantum Systems, (Lecture Notes in Physics 784)}}
  (\bibinfo{publisher}{Springer-Verlag, Heidelberg}, \bibinfo{year}{2009}).

\bibitem[{\citenamefont{Nielsen and Chuang}(2010)}]{Nielsen}
\bibinfo{author}{\bibfnamefont{M.~A.} \bibnamefont{Nielsen}} \bibnamefont{and}
  \bibinfo{author}{\bibfnamefont{I.~L.} \bibnamefont{Chuang}},
  \emph{\bibinfo{title}{Quantum computation and quantum information}}
  (\bibinfo{publisher}{Cambridge University Press}, \bibinfo{year}{2010}).

\bibitem[{\citenamefont{Modi et~al.}(2012)\citenamefont{Modi, Brodutch, Cable,
  Paterek, and Vedral}}]{RMP2012Vedral}
\bibinfo{author}{\bibfnamefont{K.}~\bibnamefont{Modi}},
  \bibinfo{author}{\bibfnamefont{A.}~\bibnamefont{Brodutch}},
  \bibinfo{author}{\bibfnamefont{H.}~\bibnamefont{Cable}},
  \bibinfo{author}{\bibfnamefont{T.}~\bibnamefont{Paterek}}, \bibnamefont{and}
  \bibinfo{author}{\bibfnamefont{V.}~\bibnamefont{Vedral}},
  \bibinfo{journal}{Rev. Mod. Phys.} \textbf{\bibinfo{volume}{84}},
  \bibinfo{pages}{1655} (\bibinfo{year}{2012}).

\bibitem[{\citenamefont{Maxwell}(1911)}]{MaxwellD}
\bibinfo{author}{\bibfnamefont{J.~C.} \bibnamefont{Maxwell}},
  \emph{\bibinfo{title}{Life and Scientific Work of Peter Guthrie Tait}}
  (\bibinfo{publisher}{edited by C. G. Knott (Cambridge University Press,
  London)}, \bibinfo{year}{1911}), p. \bibinfo{pages}{213}.

\bibitem[{\citenamefont{Szil\'{a}rd}(1929)}]{Z.Phys.53.840}
\bibinfo{author}{\bibfnamefont{L.}~\bibnamefont{Szil\'{a}rd}},
  \bibinfo{journal}{Z. Phys} \textbf{\bibinfo{volume}{53}},
  \bibinfo{pages}{840} (\bibinfo{year}{1929}).

\bibitem[{\citenamefont{Maruyama et~al.}(2009)\citenamefont{Maruyama, Nori, and
  Vedral}}]{RMP2009MaxwellD}
\bibinfo{author}{\bibfnamefont{K.}~\bibnamefont{Maruyama}},
  \bibinfo{author}{\bibfnamefont{F.}~\bibnamefont{Nori}}, \bibnamefont{and}
  \bibinfo{author}{\bibfnamefont{V.}~\bibnamefont{Vedral}},
  \bibinfo{journal}{Rev. Mod. Phys.} \textbf{\bibinfo{volume}{81}},
  \bibinfo{pages}{1} (\bibinfo{year}{2009}).

\bibitem[{\citenamefont{Zurek}(2003)}]{discorddemons}
\bibinfo{author}{\bibfnamefont{W.~H.} \bibnamefont{Zurek}},
  \bibinfo{journal}{Phys. Rev. A} \textbf{\bibinfo{volume}{67}},
  \bibinfo{pages}{012320} (\bibinfo{year}{2003}).

\bibitem[{\citenamefont{Mohammady and Anders}(2017)}]{NJP2017Szilard}
\bibinfo{author}{\bibfnamefont{M.~H.} \bibnamefont{Mohammady}}
  \bibnamefont{and} \bibinfo{author}{\bibfnamefont{J.}~\bibnamefont{Anders}},
  \bibinfo{journal}{New J. Phys.} \textbf{\bibinfo{volume}{19}},
  \bibinfo{pages}{113026} (\bibinfo{year}{2017}).

\bibitem[{\citenamefont{Elouard et~al.}(2017)\citenamefont{Elouard,
  Herrera-Mart\'{\i}, Huard, and Auff\`eves}}]{PRL2017Demon}
\bibinfo{author}{\bibfnamefont{C.}~\bibnamefont{Elouard}},
  \bibinfo{author}{\bibfnamefont{D.}~\bibnamefont{Herrera-Mart\'{\i}}},
  \bibinfo{author}{\bibfnamefont{B.}~\bibnamefont{Huard}}, \bibnamefont{and}
  \bibinfo{author}{\bibfnamefont{A.}~\bibnamefont{Auff\`eves}},
  \bibinfo{journal}{Phys. Rev. Lett.} \textbf{\bibinfo{volume}{118}},
  \bibinfo{pages}{260603} (\bibinfo{year}{2017}).

\bibitem[{\citenamefont{S\'anchez et~al.}(2019)\citenamefont{S\'anchez,
  Splettstoesser, and Whitney}}]{PRL2019NonequilibriumDemon}
\bibinfo{author}{\bibfnamefont{R.}~\bibnamefont{S\'anchez}},
  \bibinfo{author}{\bibfnamefont{J.}~\bibnamefont{Splettstoesser}},
  \bibnamefont{and} \bibinfo{author}{\bibfnamefont{R.~S.}
  \bibnamefont{Whitney}}, \bibinfo{journal}{Phys. Rev. Lett.}
  \textbf{\bibinfo{volume}{123}}, \bibinfo{pages}{216801}
  (\bibinfo{year}{2019}).

\bibitem[{\citenamefont{Beyer et~al.}(2019)\citenamefont{Beyer, Luoma, and
  Strunz}}]{PRL2019}
\bibinfo{author}{\bibfnamefont{K.}~\bibnamefont{Beyer}},
  \bibinfo{author}{\bibfnamefont{K.}~\bibnamefont{Luoma}}, \bibnamefont{and}
  \bibinfo{author}{\bibfnamefont{W.~T.} \bibnamefont{Strunz}},
  \bibinfo{journal}{Phys. Rev. Lett.} \textbf{\bibinfo{volume}{123}},
  \bibinfo{pages}{250606} (\bibinfo{year}{2019}).

\bibitem[{\citenamefont{Ji et~al.}(2022)\citenamefont{Ji, Chai, Wang, Guo,
  Rong, Shi, Ren, Wang, and Du}}]{PRL2022Ren}
\bibinfo{author}{\bibfnamefont{W.}~\bibnamefont{Ji}},
  \bibinfo{author}{\bibfnamefont{Z.}~\bibnamefont{Chai}},
  \bibinfo{author}{\bibfnamefont{M.}~\bibnamefont{Wang}},
  \bibinfo{author}{\bibfnamefont{Y.}~\bibnamefont{Guo}},
  \bibinfo{author}{\bibfnamefont{X.}~\bibnamefont{Rong}},
  \bibinfo{author}{\bibfnamefont{F.}~\bibnamefont{Shi}},
  \bibinfo{author}{\bibfnamefont{C.}~\bibnamefont{Ren}},
  \bibinfo{author}{\bibfnamefont{Y.}~\bibnamefont{Wang}}, \bibnamefont{and}
  \bibinfo{author}{\bibfnamefont{J.}~\bibnamefont{Du}}, \bibinfo{journal}{Phys.
  Rev. Lett.} \textbf{\bibinfo{volume}{128}}, \bibinfo{pages}{090602}
  (\bibinfo{year}{2022}).

\bibitem[{\citenamefont{Kim et~al.}(2011)\citenamefont{Kim, Sagawa,
  De~Liberato, and Ueda}}]{PRL2011QSzilard}
\bibinfo{author}{\bibfnamefont{S.~W.} \bibnamefont{Kim}},
  \bibinfo{author}{\bibfnamefont{T.}~\bibnamefont{Sagawa}},
  \bibinfo{author}{\bibfnamefont{S.}~\bibnamefont{De~Liberato}},
  \bibnamefont{and} \bibinfo{author}{\bibfnamefont{M.}~\bibnamefont{Ueda}},
  \bibinfo{journal}{Phys. Rev. Lett.} \textbf{\bibinfo{volume}{106}},
  \bibinfo{pages}{070401} (\bibinfo{year}{2011}).

\bibitem[{\citenamefont{Park et~al.}(2013)\citenamefont{Park, Kim, Sagawa, and
  Kim}}]{PRL2013Engine}
\bibinfo{author}{\bibfnamefont{J.~J.} \bibnamefont{Park}},
  \bibinfo{author}{\bibfnamefont{K.-H.} \bibnamefont{Kim}},
  \bibinfo{author}{\bibfnamefont{T.}~\bibnamefont{Sagawa}}, \bibnamefont{and}
  \bibinfo{author}{\bibfnamefont{S.~W.} \bibnamefont{Kim}},
  \bibinfo{journal}{Phys. Rev. Lett.} \textbf{\bibinfo{volume}{111}},
  \bibinfo{pages}{230402} (\bibinfo{year}{2013}).

\bibitem[{\citenamefont{Faist et~al.}(2015)\citenamefont{Faist, Dupuis,
  Oppenheim, and Renner}}]{NC2015}
\bibinfo{author}{\bibfnamefont{P.}~\bibnamefont{Faist}},
  \bibinfo{author}{\bibfnamefont{F.}~\bibnamefont{Dupuis}},
  \bibinfo{author}{\bibfnamefont{J.}~\bibnamefont{Oppenheim}},
  \bibnamefont{and} \bibinfo{author}{\bibfnamefont{R.}~\bibnamefont{Renner}},
  \bibinfo{journal}{Nat. Commun.} \textbf{\bibinfo{volume}{6}},
  \bibinfo{pages}{7669} (\bibinfo{year}{2015}).

\bibitem[{\citenamefont{Seah et~al.}(2020)\citenamefont{Seah, Nimmrichter, and
  Scarani}}]{PRL.124.100603}
\bibinfo{author}{\bibfnamefont{S.}~\bibnamefont{Seah}},
  \bibinfo{author}{\bibfnamefont{S.}~\bibnamefont{Nimmrichter}},
  \bibnamefont{and} \bibinfo{author}{\bibfnamefont{V.}~\bibnamefont{Scarani}},
  \bibinfo{journal}{Phys. Rev. Lett} \textbf{\bibinfo{volume}{124}},
  \bibinfo{pages}{100603} (\bibinfo{year}{2020}).

\bibitem[{\citenamefont{Perarnau-Llobet
  et~al.}(2015)\citenamefont{Perarnau-Llobet, Hovhannisyan, Huber, Skrzypczyk,
  Brunner, and Ac\'{\i}n}}]{PhysRevX.5.041011}
\bibinfo{author}{\bibfnamefont{M.}~\bibnamefont{Perarnau-Llobet}},
  \bibinfo{author}{\bibfnamefont{K.~V.} \bibnamefont{Hovhannisyan}},
  \bibinfo{author}{\bibfnamefont{M.}~\bibnamefont{Huber}},
  \bibinfo{author}{\bibfnamefont{P.}~\bibnamefont{Skrzypczyk}},
  \bibinfo{author}{\bibfnamefont{N.}~\bibnamefont{Brunner}}, \bibnamefont{and}
  \bibinfo{author}{\bibfnamefont{A.}~\bibnamefont{Ac\'{\i}n}},
  \bibinfo{journal}{Phys. Rev. X} \textbf{\bibinfo{volume}{5}},
  \bibinfo{pages}{041011} (\bibinfo{year}{2015}).

\bibitem[{\citenamefont{Mukherjee et~al.}(2016)\citenamefont{Mukherjee, Roy,
  Bhattacharya, and Banik}}]{PhysRevE.93.052140}
\bibinfo{author}{\bibfnamefont{A.}~\bibnamefont{Mukherjee}},
  \bibinfo{author}{\bibfnamefont{A.}~\bibnamefont{Roy}},
  \bibinfo{author}{\bibfnamefont{S.~S.} \bibnamefont{Bhattacharya}},
  \bibnamefont{and} \bibinfo{author}{\bibfnamefont{M.}~\bibnamefont{Banik}},
  \bibinfo{journal}{Phys. Rev. E} \textbf{\bibinfo{volume}{93}},
  \bibinfo{pages}{052140} (\bibinfo{year}{2016}).

\bibitem[{\citenamefont{Alimuddin et~al.}(2019)\citenamefont{Alimuddin, Guha,
  and Parashar}}]{PhysRevA.99.052320}
\bibinfo{author}{\bibfnamefont{M.}~\bibnamefont{Alimuddin}},
  \bibinfo{author}{\bibfnamefont{T.}~\bibnamefont{Guha}}, \bibnamefont{and}
  \bibinfo{author}{\bibfnamefont{P.}~\bibnamefont{Parashar}},
  \bibinfo{journal}{Phys. Rev. A} \textbf{\bibinfo{volume}{99}},
  \bibinfo{pages}{052320} (\bibinfo{year}{2019}).

\bibitem[{\citenamefont{Francica et~al.}(2017)\citenamefont{Francica, Goold,
  Plastina, and Paternostro}}]{francica2017daemonic}
\bibinfo{author}{\bibfnamefont{G.}~\bibnamefont{Francica}},
  \bibinfo{author}{\bibfnamefont{J.}~\bibnamefont{Goold}},
  \bibinfo{author}{\bibfnamefont{F.}~\bibnamefont{Plastina}}, \bibnamefont{and}
  \bibinfo{author}{\bibfnamefont{M.}~\bibnamefont{Paternostro}},
  \bibinfo{journal}{npj Quantum Information} \textbf{\bibinfo{volume}{3}},
  \bibinfo{pages}{1} (\bibinfo{year}{2017}).

\bibitem[{\citenamefont{Manzano et~al.}(2018)\citenamefont{Manzano, Plastina,
  and Zambrini}}]{PhysRevLett.121.120602}
\bibinfo{author}{\bibfnamefont{G.}~\bibnamefont{Manzano}},
  \bibinfo{author}{\bibfnamefont{F.}~\bibnamefont{Plastina}}, \bibnamefont{and}
  \bibinfo{author}{\bibfnamefont{R.}~\bibnamefont{Zambrini}},
  \bibinfo{journal}{Phys. Rev. Lett.} \textbf{\bibinfo{volume}{121}},
  \bibinfo{pages}{120602} (\bibinfo{year}{2018}).

\bibitem[{\citenamefont{Morris et~al.}(2019)\citenamefont{Morris, Lami, and
  Adesso}}]{PhysRevLett.122.130601}
\bibinfo{author}{\bibfnamefont{B.}~\bibnamefont{Morris}},
  \bibinfo{author}{\bibfnamefont{L.}~\bibnamefont{Lami}}, \bibnamefont{and}
  \bibinfo{author}{\bibfnamefont{G.}~\bibnamefont{Adesso}},
  \bibinfo{journal}{Phys. Rev. Lett.} \textbf{\bibinfo{volume}{122}},
  \bibinfo{pages}{130601} (\bibinfo{year}{2019}).

\bibitem[{\citenamefont{Huber et~al.}(2015)\citenamefont{Huber,
  Perarnau-Llobet, Hovhannisyan, Skrzypczyk, Kl\"{o}ckl, Brunner, and
  Ac\'{\i}n}}]{Huber_2015}
\bibinfo{author}{\bibfnamefont{M.}~\bibnamefont{Huber}},
  \bibinfo{author}{\bibfnamefont{M.}~\bibnamefont{Perarnau-Llobet}},
  \bibinfo{author}{\bibfnamefont{K.~V.} \bibnamefont{Hovhannisyan}},
  \bibinfo{author}{\bibfnamefont{P.}~\bibnamefont{Skrzypczyk}},
  \bibinfo{author}{\bibfnamefont{C.}~\bibnamefont{Kl\"{o}ckl}},
  \bibinfo{author}{\bibfnamefont{N.}~\bibnamefont{Brunner}}, \bibnamefont{and}
  \bibinfo{author}{\bibfnamefont{A.}~\bibnamefont{Ac\'{\i}n}},
  \bibinfo{journal}{New J. Phys.} \textbf{\bibinfo{volume}{17}},
  \bibinfo{pages}{065008} (\bibinfo{year}{2015}).

\bibitem[{\citenamefont{Guha et~al.}(2019)\citenamefont{Guha, Alimuddin, and
  Parashar}}]{PhysRevE.100.012147}
\bibinfo{author}{\bibfnamefont{T.}~\bibnamefont{Guha}},
  \bibinfo{author}{\bibfnamefont{M.}~\bibnamefont{Alimuddin}},
  \bibnamefont{and} \bibinfo{author}{\bibfnamefont{P.}~\bibnamefont{Parashar}},
  \bibinfo{journal}{Phys. Rev. E} \textbf{\bibinfo{volume}{100}},
  \bibinfo{pages}{012147} (\bibinfo{year}{2019}).

\bibitem[{\citenamefont{Buffoni et~al.}(2019)\citenamefont{Buffoni, Solfanelli,
  Verrucchi, Cuccoli, and Campisi}}]{PRL2019MCool}
\bibinfo{author}{\bibfnamefont{L.}~\bibnamefont{Buffoni}},
  \bibinfo{author}{\bibfnamefont{A.}~\bibnamefont{Solfanelli}},
  \bibinfo{author}{\bibfnamefont{P.}~\bibnamefont{Verrucchi}},
  \bibinfo{author}{\bibfnamefont{A.}~\bibnamefont{Cuccoli}}, \bibnamefont{and}
  \bibinfo{author}{\bibfnamefont{M.}~\bibnamefont{Campisi}},
  \bibinfo{journal}{Phys. Rev. Lett.} \textbf{\bibinfo{volume}{122}},
  \bibinfo{pages}{070603} (\bibinfo{year}{2019}).

\bibitem[{\citenamefont{Jussiau et~al.}(2023)\citenamefont{Jussiau, Bresque,
  Auff\`eves, Murch, and Jordan}}]{PhysRevResearch.5.033122}
\bibinfo{author}{\bibfnamefont{E.}~\bibnamefont{Jussiau}},
  \bibinfo{author}{\bibfnamefont{L.}~\bibnamefont{Bresque}},
  \bibinfo{author}{\bibfnamefont{A.}~\bibnamefont{Auff\`eves}},
  \bibinfo{author}{\bibfnamefont{K.~W.} \bibnamefont{Murch}}, \bibnamefont{and}
  \bibinfo{author}{\bibfnamefont{A.~N.} \bibnamefont{Jordan}},
  \bibinfo{journal}{Phys. Rev. Res.} \textbf{\bibinfo{volume}{5}},
  \bibinfo{pages}{033122} (\bibinfo{year}{2023}).

\bibitem[{\citenamefont{Purkait and Biswas}(2023)}]{PhysRevE.107.054110}
\bibinfo{author}{\bibfnamefont{C.}~\bibnamefont{Purkait}} \bibnamefont{and}
  \bibinfo{author}{\bibfnamefont{A.}~\bibnamefont{Biswas}},
  \bibinfo{journal}{Phys. Rev. E} \textbf{\bibinfo{volume}{107}},
  \bibinfo{pages}{054110} (\bibinfo{year}{2023}).

\bibitem[{\citenamefont{Sagawa and Ueda}(2009)}]{PRL2009cost}
\bibinfo{author}{\bibfnamefont{T.}~\bibnamefont{Sagawa}} \bibnamefont{and}
  \bibinfo{author}{\bibfnamefont{M.}~\bibnamefont{Ueda}},
  \bibinfo{journal}{Phys. Rev. Lett.} \textbf{\bibinfo{volume}{102}},
  \bibinfo{pages}{250602} (\bibinfo{year}{2009}).

\bibitem[{\citenamefont{Geva and Kosloff}(1992)}]{geva1992quantum}
\bibinfo{author}{\bibfnamefont{E.}~\bibnamefont{Geva}} \bibnamefont{and}
  \bibinfo{author}{\bibfnamefont{R.}~\bibnamefont{Kosloff}},
  \bibinfo{journal}{J. Chem. Phys.} \textbf{\bibinfo{volume}{96}},
  \bibinfo{pages}{3054} (\bibinfo{year}{1992}).

\bibitem[{\citenamefont{Kieu}(2004)}]{kieu2004second}
\bibinfo{author}{\bibfnamefont{T.~D.} \bibnamefont{Kieu}},
  \bibinfo{journal}{Phys. Rev. Lett.} \textbf{\bibinfo{volume}{93}},
  \bibinfo{pages}{140403} (\bibinfo{year}{2004}).

\bibitem[{\citenamefont{Quan et~al.}(2007)\citenamefont{Quan, Liu, Sun, and
  Nori}}]{quan2007quantum}
\bibinfo{author}{\bibfnamefont{H.~T.} \bibnamefont{Quan}},
  \bibinfo{author}{\bibfnamefont{Y.-X.} \bibnamefont{Liu}},
  \bibinfo{author}{\bibfnamefont{C.~P.} \bibnamefont{Sun}}, \bibnamefont{and}
  \bibinfo{author}{\bibfnamefont{F.}~\bibnamefont{Nori}},
  \bibinfo{journal}{Phys. Rev. E} \textbf{\bibinfo{volume}{76}},
  \bibinfo{pages}{031105} (\bibinfo{year}{2007}).

\bibitem[{\citenamefont{Thomas and Johal}(2011)}]{thomas2011coupled}
\bibinfo{author}{\bibfnamefont{G.}~\bibnamefont{Thomas}} \bibnamefont{and}
  \bibinfo{author}{\bibfnamefont{R.~S.} \bibnamefont{Johal}},
  \bibinfo{journal}{Phys. Rev. E} \textbf{\bibinfo{volume}{83}},
  \bibinfo{pages}{031135} (\bibinfo{year}{2011}).

\bibitem[{\citenamefont{Du and Zhang}(2018)}]{Du_2018}
\bibinfo{author}{\bibfnamefont{J.-Y.} \bibnamefont{Du}} \bibnamefont{and}
  \bibinfo{author}{\bibfnamefont{F.-L.} \bibnamefont{Zhang}},
  \bibinfo{journal}{New J. Phys.} \textbf{\bibinfo{volume}{20}},
  \bibinfo{pages}{063005} (\bibinfo{year}{2018}).

\bibitem[{\citenamefont{Baumgratz et~al.}(2014)\citenamefont{Baumgratz, Cramer,
  and Plenio}}]{PhysRevLett.113.140401}
\bibinfo{author}{\bibfnamefont{T.}~\bibnamefont{Baumgratz}},
  \bibinfo{author}{\bibfnamefont{M.}~\bibnamefont{Cramer}}, \bibnamefont{and}
  \bibinfo{author}{\bibfnamefont{M.~B.} \bibnamefont{Plenio}},
  \bibinfo{journal}{Phys. Rev. Lett.} \textbf{\bibinfo{volume}{113}},
  \bibinfo{pages}{140401} (\bibinfo{year}{2014}).

\end{thebibliography}
	
\end{document}